# Detecting Activities of Daily Living and Routine Behaviours in Dementia Patients Living Alone Using Smart Meter Load Disaggregation

C. Chalmers, P.Fergus, C. Aday Curbelo Montanez, S.Sikdar, F.Ball and B. Kendall

*Abstract*— The emergence of an ageing population is a significant public health concern. This has led to an increase in the number of people living with progressive neurodegenerative disorders like dementia. Consequently, the strain this is places on health and social care services means providing 24-hour monitoring is not sustainable. Technological intervention is being considered, however no solution exists to non-intrusively monitor the independent living needs of patients with dementia. As a result many patients hit crisis point before intervention and support is provided. In parallel, patient care relies on feedback from informal carers about significant behavioural changes. Yet, not all people have a social support network and early intervention in dementia care is often missed. The smart meter rollout has the potential to change this. Using machine learning and signal processing techniques, a home energy supply can be disaggregated to detect which home appliances are turned on and off. This will allow Activities of Daily Living (ADLs) to be assessed, such as eating and drinking, and observed changes in routine to be detected for early intervention. The primary aim is to help reduce deterioration and enable patients to stay in their homes for longer. A Support Vector Machine (SVM) and Random Decision Forest classifier are modelled using data from three test homes. The trained models are then used to monitor two patients with dementia during a six-month clinical trial undertaken in partnership with Mersey Care NHS Foundation Trust. In the case of load disaggregation for appliance detection, the SVM achieved (AUC=0.86074, Sen=0.756 and Spec=0.92838). While the Decision Forest achieved (AUC=0.9429, Sen=0.9634 and Spec=0.9634). ADLs are also analysed to identify the behavioural patterns of the occupant while detecting alterations in routine. The results show that the approach is highly sensitive in identifying behavioural routines and detecting anomalies in patient behaviour.

## I. INTRODUCTION

Assistive technology covers a wide range of tools and techniques to support independent living in domiciliary settings [1]. Particular interest, in recent years, has focused on monitoring technologies for early intervention services and out-patient condition management [1]. Typical solutions include physical aids and remote surveillance. Both approaches are designed to help patients perform daily tasks and support their healthcare needs, automatically alerting healthcare staff and relatives as and when required [2]. Many existing approaches depend on multi-sensor deployment in homes and on patients themselves [2]. These include motion sensors, cameras, fall detectors and communication hubs, which all need patient interaction.

In healthcare settings, these assistive technologies are also proprietary and often tailored to specific application scenarios. They are rarely personalised to meet the specific needs of a patient and rarely identify normal and abnormal trends in behaviour. This is due to the fact assisted living technologies have limited utility in unobtrusively detecting Activities of Daily Living (ADLs). Consequently, adherence in both patients and healthcare professionals is low resulting in most telehealth solutions never being used [3]. The smart meter rollout has the potential to change this. In this paper, we propose a system that allows a person with dementia to go about their day-to-day life and retain their independence, while monitoring their habits to provide greater piece of mind for the person's family and carers. This system does not require direct interaction from the patient. Instead, data is collected about an individual's habits and routines through their normal interaction with electrical devices, i.e. putting the kettle on or heating a meal in the microwave. This allows the patient's ADLs to be assessed; any changes observed will trigger an early intervention to help reduce excessive deterioration in some cases; enabling the patient to stay in their home for longer.

The approach is foundational and builds on worldwide, government financed, Smart Meter rollout programs - 55% of installations are expected to be complete by the end of 2020 [4]. The Smart Meter infrastructure provides a unique opportunity to deliver healthcare solutions to patients living alone with dementia. This will provide support and care packages that are tailored to the individual needs of a patient, ensuring a much closer relationship between patients and those who care for them. The approach is novel and significantly superior to any other telehealth solution proposed worldwide in dementia care. The results are significant and we argue in this paper that the approach warrants wider discussion with national healthcare services and relevant governmental departments.

The remainder of the paper is structured as follows. A background discussion on current AAL solutions and their associated limitations is introduced in Section 2. Section 3 discusses smart meters, the concept of Non-Intrusive Load Monitoring (NILM) and the challenges involved in processing smart meter data for the purpose of load disaggregation. Section 4 describes the proposed approach and presents the results obtained from a clinical trial conducted in partnership with Mersey Care NHS Foundation Trust in the UK. The results are discussed in Section 5 before the paper is concluded and future work is presented in Section 6.

## II. CURRENT AMBIENT ASSISTIVE LIVING TECHNOLOGIES

Several clinical trials in Ambient Assistive Living (AAL) have been conducted to assess the feasibility for using technology in healthcare practices [5], [6]. AAL utilises technology (i.e. sensors, computing etc.) within and across different domains (i.e. computer science, engineering, medicine and social sciences) to identify human activities and provide medical insights; commonly referred to as telehealth.

Telehealth has undoubtedly helped patients to live independently which, according to Mordor Intelligence, will be worth $66 billion by the end of 2021[1]. However, there are numerous examples where telehealth solutions were anticipated to transform healthcare but failed to deliver. In many instances their use in healthcare increases overall care plan costs by 10%, yet a recent study showed that many solutions only provide minimal gains in a patient's quality of life [7]. More specifically, the cost of each Quality Adjusted Life Year (QALY) achieved between intervention and non-intervention groups equates to 0.012 QALYs (only a few additional days of quality health). This is below the cost-effective threshold recommended by The National Institute for Health and Care Excellence (NICE) [7].

### A. Current Smart Home Solutions

AAL technology provides two main types of monitoring: preventative and responsive. The preventative model minimises patient risks using ADLs, by supporting tasks such as taking medications, eating and drinking. Responsive models on the other hand react to events like falls, alarms and patients leaving their home. Often responsive technologies only work well when predefined protocols are defined.

The medical profession is interested in both approaches and their application in delivering home healthcare services. The rationale being that if smart homes can be used to deliver healthcare services, then it would be possible to monitor patients and detect relapse indicators to support early intervention practices [8]. This model fits well with patients and carers alike as over 80% would prefer to stay in their own home in the later years of life[2].

The fact is, monitoring patients remotely provides new and interesting ways of supporting patients during their dementia journey. Currently, there are several ways to achieve this. Table 1 provides a brief summary of common solutions [9].

TABLE 1

SENSORS DEPLOYED IN A SMART ENVIRONMENT

| Sensor Type | Measurement | Limitations |
|---|---|---|
| Passive Infrared Motion Sensor (PIR) | Movement around the living environment. | Multiple sensors are required. Typically one for each room in a persons living environment. PIR solutions often fail to detect key ADLS, as they can only verify location and not the occupant's activity. Sensors can have poor battery life, which requires ongoing maintenance and accurate detection of failing batteries. |
| Radio Frequency Identification (RFID) | Movement around the living environment. | Multiple sensors are required, which are distributed throughout the living environment. RFID often suffers from reduced accuracy due to interference from neighbouring sensors [10]. It is common for RFID solutions to experience contact sensing difficulties. For example, when a sensor is within the range of an antenna but is not detected. |
| Pressure Sensors / Smart Tiles | Detects the presence of pressure on multiple items such as flooring, mats, beds and chairs. | Often inaccurate (sensing motion not presence) [11]. Equipment positioning often requires important consideration to obtain the best results. In addition, they are often used in conjunction with other sensors [12]. |
| Magnetic Switches | Detection of door / cabinet opening and closing. | Multiple sensors required. Switches can be wired or wireless and are often used in conjunction with other sensors. |
| Cameras | Tracks activity within the living environment. | Often considered unacceptable due to legal, privacy and ethical issues. Additionally, the deployment of camera technology within the living environment is both expensive and intrusive [13]. |
| Microphone | Used to record and identify particular noises within the home. | Microphones can be deployed throughout the living environment and can be used to identify significant sounds. Noises can be utilised for the detection of ADLS or identifying if the patient is in trouble. |
| Physical Alarms | Devices, which are worn by the patient and can be triggered in the event of an emergency. | Systems that are solely reliant on a person's interaction to function pose many safety concerns. Dementia patients in particular may forget to activate the device or fail to identify if they are at risk. |

### B. Limitations

There are no agreed standards for AAL and concerns surrounding high costs and complex installations impede their adoption [14]. In many instances, they are too rigid and therefore simply fail to meet the unique requirements needed by patients and their home environments to facilitate independent living [15]. State-of-the-art products are propriety configurations that do not interoperate with other solutions

---
[1] https://www.forbes.com/sites/quora/2018/07/31/what-are-the-latest-trends-in-telemedicine-in-2018/#5db0b96e6b9e
[2] http://www.raeng.org.uk/publications/reports/designing-cost-effective-care-for-older-people

and services, or facilitate Early Intervention Practice (EIP) and unobtrusive ADL detection.

They are often considered intrusive and incapable of being personalised to the individual. There are no mechanisms to learn the unique characteristics of patients and their conditions; this limits their effectiveness in most home healthcare services. Consequently, large scale adoption within National Health Service (NHS) trusts, councils and social services is not feasible.

### III. APPROACH

The proposed solution utilises the smart meter infrastructure to deliver dementia care support. This system does not require direct interaction from the patient. Instead, data is collected about an individual's habits and routines through their normal interaction with electrical devices. The system addresses three fundamental limitations associated with existing AAL technologies: unobtrusive ADL detection, EIP support, and per patient personalisation. This is achieved using data obtained from the smart metering infrastructure to better understand an individual's habits and routines through interactions with electrical devices. This allows patients ADLs to be assessed; changes in normal routines are used trigger an early intervention.

Smart meters generate large amounts of energy usage data [16]. Utilising this data facilitates applications outside of typical energy management scenarios. By utilising an integrated system of smart meters, communication networks and data management systems, two-way communications between energy suppliers and consumers is possible. This section discusses the fundamental components that make up the infrastructure and the functionality provided before the details of the proposed solution are presented.

#### A. Smart Meter Infrastructure

The main objective of the smart grid is to balance energy load effectively through a network of connected smart meters [17]. According to the International Energy Agency (IEA), smart grid technologies are essential to meet future energy requirements [18]. The IEA expects worldwide energy demands to increase at an annual rate of 2.2%, eventually doubling the global energy demand by 2040 [19]. In the smart grid architecture, all energy consumption data is acquired directly from individual smart meters. This data is stored, managed and analysed in the Meter Data Management System (MDMS) [20]. Figure 1 provides an overview of a typical MDMS.

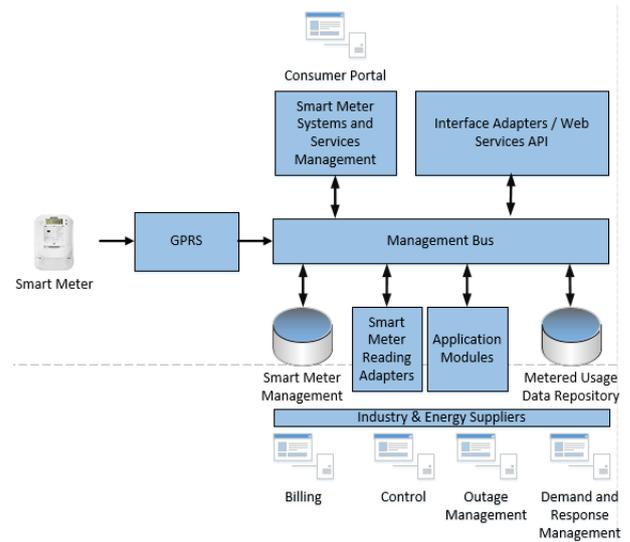

**Fig. 1**: Typical MDMS Overview

The MDMS sits within the data and communications layer of the AMI and is a scalable software platform that provides data analytic services for AMI applications. Applications include, managing metered consumption data, outage management, demand and response, remote connect/disconnect, smart meter events and billing [21]. The information generated can be shared with consumers, partners, market operators and regulators.

Data processing within the AMI is a significant challenge. Smart meters in the UK collect and transmit energy usage information to the MDMS every 30 minutes [22]. Lower sampling rates are possible, but managing increases in data throughput has a significant financial impact on utility companies. Table 2 provides a summary of typical data sources found in the smart grid infrastructure.

TABLE 2

SMART GRID DATA SOURCES

| Data Type | Technology | Description |
|---|---|---|
| AMI | Smart Meters | Consumption data that is generated from smart meters at a predefined frequency. |
| Distribution and Automation | Grid Equipment | The distribution and automation system, that collects data from the various sensors distributed throughout the entire grid. These sensors can generate up to 30 readings per second per sensor. |
| Third - Party | External Data Sets | The integration of 3rd party data, such as demand and response. |
| Asset Management | OS / Firmware | Communication between the MDMS and smart technologies. |

The collection and accessibility of data does not provide any real value without software tools and expertise to exploit it. Consequently, data science has become a major focus for smart grid research [23]. The primary interest is to extract meaningful information from selected datasets for decision-making and service provisioning directly on top of the smart meter infrastructure.

## B. Load Disaggregation

Smart meter data opens up new opportunities to exploit smart meter energy readings to identify individual electrical devices and the habitual usage patterns of people in their home. Appliance Load Monitoring (ALM) for example is already being analysed to identify different appliance types. [24]. ALM is divided into two categories: Non-Intrusive Load Monitoring (NILM) and Intrusive Load Monitoring (ILM). NILM is a single point sensor, such as a smart meter. In contrast, ILM is a distributed sensing method that uses multiple sensors – one for each electrical device being monitored [25].

ILM is a more accurate method given that readings are obtained directly from the device [26]. However, this approach has a financial implication and is complex to set up. Within an ILM platform sensors can be moved between different devices and this can skew identification and classification results. While NILM is regarded as less accurate and more challenging, it mitigates this issue as appliances are identified from aggregated energy readings obtained for the whole property [27]. This nonetheless requires algorithms to identify appliance power signatures from aggregated load readings that correlate with individual appliance states. In the approach presented in this paper data is obtained from the smart meter directly and is defined as:

$$P_t = p1_t + p2_t + \ldots + pn_t \quad (1)$$

where $p$ is the power consumption of individual devices that contribute to the total aggregated measurement, and $n$ is the total number of devices within the time period $t$.

In a typical NILM environment, simple hardware is deployed and complex software tools are used to identify different appliances. NILM comprises four key stages, *data acquisition*, *event detection*, *appliance feature extraction* and *device classification*. The overall accuracy of appliance detection in NILM is significantly influenced by the sample rate [28]. Low sampling increases the number of errors in device identification due to event triggers being overlooked. Therefore, there are two common sampling rates used: those greater than 60Hz, and those less than 60Hz. Alternative configurations can be adopted where more advanced features are required, for example, voltage, current, real power, power factor, phase angle and reactive power.

## C. Electrical Device Types

Electrical appliances run in multiple modes, alongside their normal on-off states. For example, many devices have low power requirements or standby modes. While appliances like ovens can operate using several control functions. Understanding different device categories is vital for NILM, as they provide different information on electrical usage characteristics. Device categories include:

- Type 1 devices: operate in two states either on or off. Examples include kettles, toasters and lighting. Figure 2 shows a power reading for a kettle - (a) highlights a series of devices being used in conjunction or in close succession; while (b) presents evenly distributed single device interactions.

- Type 2 devices: are known as Multi-State Devices (MSD) or finite state appliances. They operate in multiple states with more complex behaviours than simple on-off states. Devices include washing machines, dryers and dishwashers.

- Type 3 devices: are known as Continuously Variable Devices (CVD). Their power draw has no fixed state. There is no repeatability in their characteristics and as such they are problematic in NILM. Examples devices include power tools.

- Type 4 devices: are fairly new in terms of device category. These devices are active for long periods and consume electricity at a constant rate [29]. Type 4 devices are always on therefore there are no major events to detect other than small fluctuations. Such devices include smoke detectors and intruder alarms.

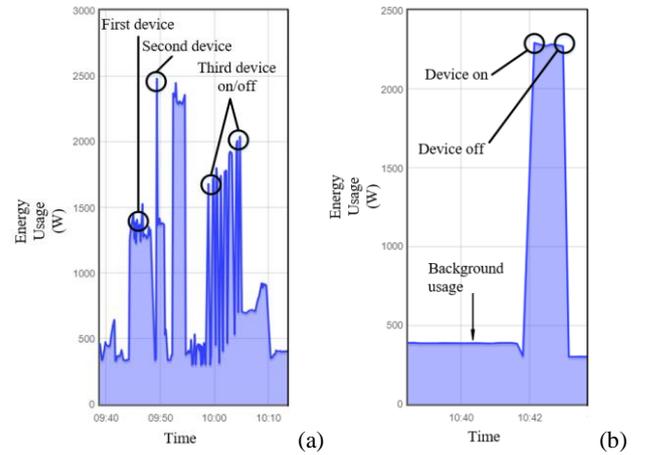

**Fig. 2:** Device Interaction Signatures

NILM appliance state detection is defined by Parson *et al.*, where *z* is the appliance state [30]:

$$Z^n = z_1^n, \ldots, z_t^n \quad (2)$$

Understanding these differences is important in any load disaggregation system, as devices are often used in combination, typically when preparing meals. This can affect the performance in classification tasks due the boundaries that exist between device classes, making them difficult to identify. The natural boundaries between different device classes are illustrated in Figure 3, use Yifan Hu clustering [31]. Specific device types are grouped based on an association, defined by their electricity consumption patterns.

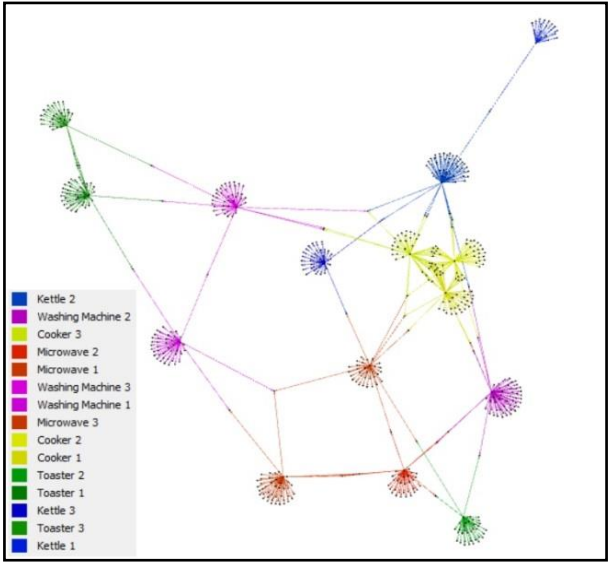

**Fig. 3:** Cluster Visualisation of Device Types 1 and 2

Yifan Hu is a force-directed algorithm [31] that calculates attraction and repulsion forces. This visually demonstrates the similarity of device types through a clustering process. The repulsion $F_r$ formula is defined as:

$$F_r = \frac{k}{d^2} \qquad (3)$$

while the attraction $F_a$ formula is expressed as:

$$F_a = -k \cdot d \qquad (4)$$

where $d$ represents the distance between the two nodes. One of the main benefits is that specific structure in the data can be calculated using information contained within the graph itself. This removes the need for domain-specific knowledge [31].

The algorithm uses the repulsive forces on one *node* from a cluster of distant *nodes*. In this case, nodes correspond to electrical device readings. These are approximated by a Barnes-Hut calculation scheme for grouping together bodies that are sufficiently nearby [31]. It uses a multilevel approach to find globally optimal layouts, and the Barnes-Hut octree technique to approximate short and long range forces [31]. Typically, this multilevel approach has three phases, as shown in Algorithm 1 [31]. The starting point is the original graph, $G_0 = G$ and $n^i = |V^i|$ are the coefficients for the number of vertices in the $i^{th}$ level graph, represented as $G^i$. $x^i$ is defined as the coordinate vector for the vertices in $V^i$. $G^i$ is represented by a symmetric matrix $G^i$, where all entries of the matrix act as the edge weights. $G_{i+1}$ to $G_i$ is the continuation operator, also represented by a matrix $P^i$, of dimension $n^i * n^{i+1}$.

---

**Algorithm 1** Three Phase Descriptors

1: **Coarsest Graph Layout:**

   **if** ($n^{i+1} <$ MinSize or $\frac{n^{i+1}}{n^i} > p$){

   ∗ $x^i \coloneqq$ random initial layout

   ∗ $x^i =$ ForceDirectedAlgorithm$(G^i, x^i, \text{tol})$

   ∗ **return** $x^i$ }

2: **The Coarsening Phase:**

   set up the $n^i \times n^{i+1}$ prolongation matrix $P^i$

   $G^{i+1} = P^{i^T} G^i P^i$

   $x^{i+1} =$ MultilevelLayout$(G^{i+1}, \text{tol})$

3: **The Prolongation and Refinement Phase:**

   prolongate to get initial layout: $x^i = P^i x^{(i+1)}$

   refinement: $x^i =$ ForceDirectedAlgorithm$(G^i, x^i, \text{tol})$

   **return** $x^i$

---

The graph displays a visualisation of raw data collected from smart meters. The boundaries between classes provide guidance on what classifiers to use (i.e. linear, quadratic or polynomial) within the same feature space. For example, overlapping data shows that a linear division of the raw data is not possible. As such, both a quadratic and axis aligned algorithms are utilised in our NILM approach.

### D. Condition Monitoring

Condition monitoring in dementia patients is challenging due to different levels of memory impairment and social disengagement. There is however a common set of features for Alzheimer's disease (and some other dementias), these include changes in physical health such as falls, Urinary Tract Infection (UTI), dehydration, sleep disorder, depression, apathy and appetite disturbances, that could potentially be detected through ongoing interactions with home appliances. The severity of each symptom differs at various stages of the disease so systems need to be adaptable to these changes, as patients progress through different stages of the illness.

Behavioural problems, such as agitation, become more pronounced in the later stages of the disease. For example, during periods of severe depression, the patient may interact less with their electrical devices, they may stay in bed for longer periods of time (insomnia or hypersomnia) or not cook meals (change in appetite). Changes in sleep behaviours and appetite are all reflected through energy usage. Such behaviours can be identified and investigated further where appropriate. Likewise, alterations in appliance usage, i.e. operating appliances during abnormal times of the day, provide key indicators that patients may be experiencing difficulties. Many of these symptoms can be detected by analysing electricity usage patterns.

Of the various cognitive screening tests used in diagnosing and staging severity of dementia, the Mini Mental State Examination (MMSE) is one that is commonly used by clinicians. After a patient's initial diagnosis, ongoing evaluation is required but systems vary throughout the country. Typically, a patient is reassessed three months after diagnosis and twice a year thereafter. Such sporadic monitoring does not provide sufficient granularity to adequately care for patients and so by consistently monitoring their ADLs through smart meter data, disease progression can be identified much earlier. Figure 4 highlights the MMSE, showing the need for changes in the monitoring techniques as the severity of the disease increases.

To achieve effective monitoring, it is important to identify significant behavioural traits within the target group. The identification of expected behaviour and potential relapse indicators aids in the selection of appropriate analytical techniques. Establishing routines (mapping appliance usage to specific observation periods) facilitates the detection of abnormal behaviour.

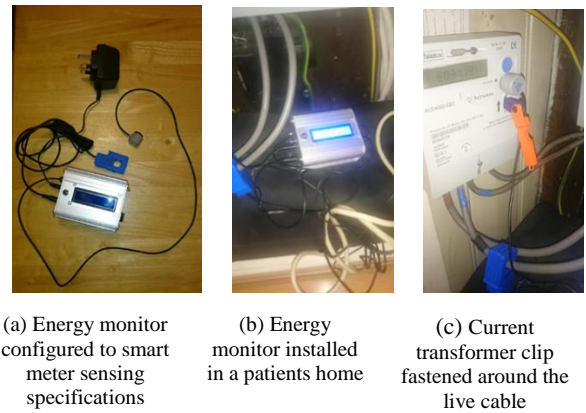

(a) Energy monitor configured to smart meter sensing specifications

(b) Energy monitor installed in a patients home

(c) Current transformer clip fastened around the live cable

**Fig. 5:** Electricity Monitor

Note that at the time of the trial none of the participants had a smart meter. Therefore, the system illustrated in 5 was configured to simulate the smart meter and an associated Consumer Access Device (CAD) to obtain 10 second data. The solution shown in (a) is connected to a home's live cable using a current sensor transformer clip (CT) shown in (b) to measure the electrical load every 10 seconds. A second Optical Pulse sensor, shown in (c), works by sensing the LED pulse output from the utility meter. The data obtained is logged to a cloud database using the patient's WIFI. This configuration, when a SMETS II smart meter is installed with a CAD, would not be required – this is only used in the case where smart meters and a CAD are not installed. Typically, data would be redirected from the CAD to the cloud database using cellular communications.

*A. System Functionality*

Once installed, the unique energy signatures are identified for each device and used to establish ADL routines. The framework operates in three specific modes; device training mode, behavioural training mode and prediction mode.

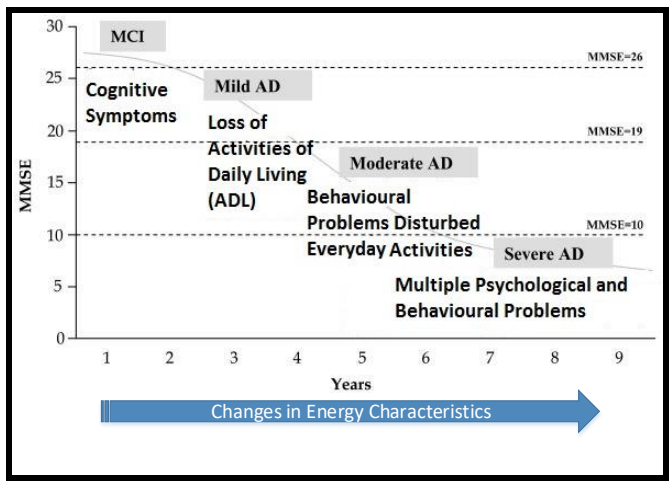

**Fig. 4.** MMSE Framework

IV. IMPLEMENTATION & CLINICAL TRAIL

This system presented in this paper has been tested in a clinical trial with two people living with dementia over a 6 month period, cared for by Mersey Care NHS Foundation Trust[3]. In line with NHS Research Ethics Committee (REC) recommendations, participants were recruited who had the capacity to consent to the study and who lived alone.

During the clinical trial an energy monitor was installed in each patient's home. This was undertaken to identify interactions with electrical appliances while establishing behavioural routines. Figure 5 provides an overview of the electricity monitoring system.

- Mode 1 (device training): power readings are obtained from the patient's smart meter and recorded in a data store. Readings are used to train the system to identify device signatures from aggregated load readings. The training process achieves this using machine learning classifiers.

- Mode 2 (behavioural training): data features are extracted to identify normal and abnormal patterns in behaviour. The features allow the system to recognise the daily routines performed by patients, including their particular habits and behavioural trends.

- Mode 3 (prediction mode): the detection of both normal and abnormal patient behaviours is conducted in real-time. The framework uses web services to facilitate machine-to-machine communication using a collection of open protocols, API's and standards. During this process, the monitoring application interfaces with web services to receive real-time monitoring alerts about the patient's wellbeing. Figure 6 shows an electric kettle being detected in a patient's home during the clinical trial. Here

---

[3] http://www.merseycare.nhs.uk/knowledge-hub/mental-health-articles/smart-meters-study/

the real-time energy readings are shown along with the device classification.

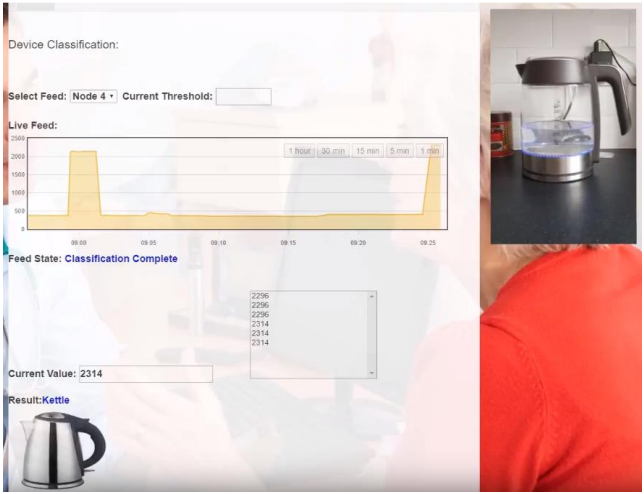

**Fig. 6**: Real-time Device Classification

The completed end-to-end system is shown in Figure 7.

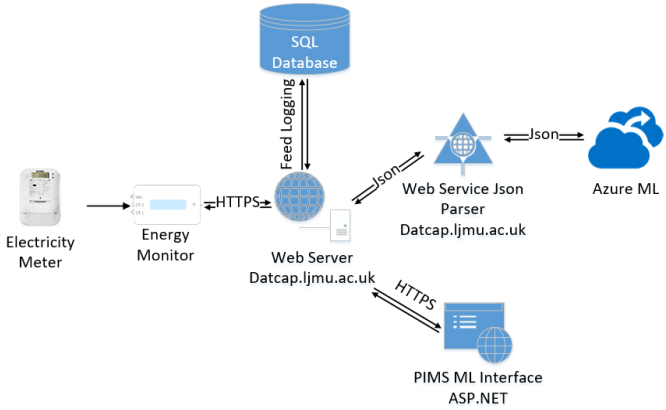

**Fig. 7:** System Framework Used in Clinical Trail

### B. Data Collection

The training dataset for device classification was constructed using energy monitors installed in three separate homes (not part of the clinical trial). Using device specific signal signatures, five distinct classes are generated; kettle, toaster, microwave, washing machine and electric oven (appliances sufficient enough to detect ADLs). The data set contains 25 individual samples from each property totalling 75 for each device class. As the aggregated readings are sampled at each 10 second interval there are 450 individual data points for each class and 2250 for the entire dataset. Empirically this provided sufficient data points to accurately detect each of the five appliances.

### C. Load Disaggregation Fast Fourier Transform

The time series readings obtained during data collected are filtered and transformed before data processing tasks are performed. A highpass filter is applied to background noise below 300 watts – signals below this threshold typically represent type 4 electrical appliances. Using an adapted Fast Fourier Transform (FFT) the filtered time series signals are converted to the frequency domain and the range of frequency values for different appliances of the same class are extracted and used to train the machine learning algorithms for appliance classification tasks. This is an import aspect of the approach as their will be variance within appliance classes (different wattage values), machine learning allows us to use these variant values to generalise across different manufactured devices, i.e. different types of kettle or microwave. Using an FFT approach allows us to detect any number of devices that are simultaneously on as they occupy distinctly different frequency groups.

### D. Data Processing

All data is normalisation to eliminate bias due to differences in data scaling, Min-Max is used in this study, where data is scaled between 0-1 as defined by:

$$x' = \frac{(x - min(x))}{(max(x) - min(x))} \quad (6)$$

Two feature selection techniques are evaluated in this study - Fisher Linear Discriminant Analysis (FLDA) and Spearman Correlation (SC) [32] [33]. FLDA works well when the variance between groups is similar and data is normally distributed. FLDA obtains a linear combination of features that determines the direction the classes are separated most accurately. When considering various classes, the distance between the means is calculated to find a linear combination of features:

$$f(y) = W^T X + \alpha \quad (7)$$

Where $\alpha$ is the bias, $W$ is calculated using Fishers LDA, and $X$ is the training data without class labels. For a multiclass approach, a one-verse-all method is employed based on [34] and defined as:

$$\sum_W = \sum_{i=1}^{C} \sum_{x_k \in X_i} (x_k - \mu_i)(x_k - \mu_i)^t \quad (8)$$

Where $C$ refers to the quantity of classes, $X_i$ the set of points in class $i$, $\mu_i$ the mean of class $i$, and $X_k$ the $k^{th}$ point of $X_i$. The subsequent scatter matrix is the correlation of class means [35] and is defined as:

$$\sum_B = \frac{1}{C} \sum_{i=1}^{C} (\mu_i - \mu)(\mu_i - \mu)^t \quad (9)$$

Where $C\text{-}1$ is the principal eigenvalue, $N_i$ the values that belong to Class $i$, $\mu_i$ the mean of Class $i$, and $\mu$ the overall mean. Spearman Correlation utilises a non-parametric test to ascertain the statistical dependence between observational stochastic sequences. It assesses the relationship among the

sequences in which the coefficient can be depicted using a monotonic function:

$$\rho = 1 - \frac{6 \sum d_i^2}{n(n^2 - 1)} \quad (10)$$

Where $\rho$ denotes the Spearman rank correlation coefficient, $d$ is the difference between the sequences, and $n$ is the number of sequences. Here a subset of features is generated with the highest degree of predictive power. Consequently, each column is scored and later utilised to build the predictive model.

Figure 8 highlights the distribution of the training data using Quantile – Quantile Q – Q plot. As shown, none of the device signatures follow a normal distribution. Understanding the data distribution informs both the feature selection techniques required and the classifiers that are likely to provide the best performance.

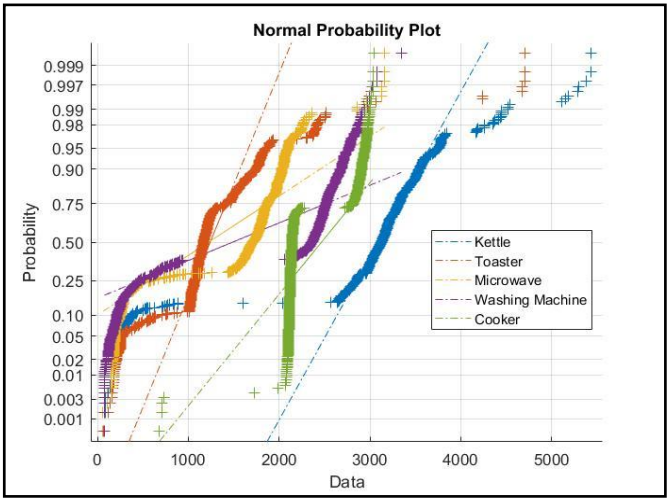

**Fig. 8:** Distribution of Device Training

### E. Machine Learning as a Cloud Service

Machine learning classifiers are utilised to generalise device detection from aggregate load. Two classifiers are considered in this study - a Multiclass Decision Forest and a Two Class SVM. The SVM is adapted to support a multiclass operation through a one verse all approach. The SVM is trained using a distribution-free learning process [36] and is defined as:

$$D = \{(X_i, Y_i) \in X \times Y\}, i = 1, l \quad (11)$$

Where $l$ is the number of training data pairs equal to the size of the training data set $D$ and $Y_i$ the desired target output [37].

A Multiclass Decision Forest is built from multiple decision trees and a '*voting*' function to determine the most popular class. Each tree in a decision forest outputs a non-normalised frequency histogram of labels. The aggregation process sums these histograms and normalises the result to get the probabilities for each label. Decision forests are both scalable and flexible and can be adjusted for each classification task. For simple problems, they enable the user to select node parameters manually, while in more complex scenarios, tree structures and parameters can be learned automatically using training data.

Multiclass Decision Forests support several configurable parameters. The Randomness parameter is introduced to trees during the training phase using bootstrap aggregation or bagging [38]. Bagging belongs to an ensemble method, which combines multiple predictions to generate an accurate model, such that:

$$\hat{f} = \frac{1}{B} \sum_{b=1}^{B} f_b(x') \quad (12)$$

Where, $f_b$ is the decision tree, $B$ is the number of times bagging takes place and $x$ is the training set. Each tree is trained on a new sample, which is generated by randomly sampling the training data; essentially each tree utilises a different training subset. Each output (prediction) is combined to generate an accurate prediction by majority voting or by averaging the results in order to obtain the best outcome. Bagging increases training speed and efficiency, while decreasing model variance, and managing bias.

### F. Vectors for Behavioural Analysis

Individual device detections from classifiers are stored as feature vectors for behavioural analysis. The predicted class is assigned a unique device ID for each appliance class detected.

The observation window can be adjusted to a patient needs and the medical condition under investigation to identify abnormal behaviours. This allows the system to construct a personalised representation of the patient, as device usage can be assigned to specific observation windows. This approach uncovers routines and alterations in behavioural. The system uses 7 distinct observation windows in a 24-hour period to achieve this as highlighted in figure 9.

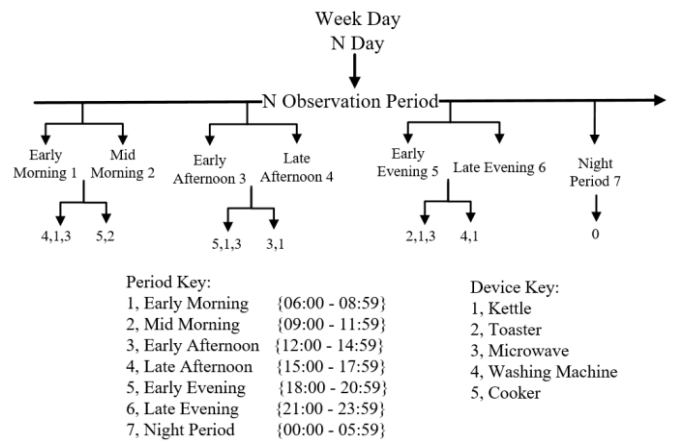

**Fig. 9:** Distribution of Device Training

Device classifications are assigned to a specific observation window, depending on the time of day. These are then used to map aspects of the patient's routine.

To identify the degree of correlation between appliance usage and the hour of day Sankey Diagram are used as illustrated in Figure 10. Sankey diagrams provide quantitative information relating to flows, including relationships and transformations [39]. To identify anomalies in device usage the anomaly detection the Z-score technique is employed. Here the Z-score describes a data point in terms of its relationship to the mean and standard deviation of a group of points.

## V. EVALUATION AND DISCUSSION

In this evaluation the performance of the Decision Forest and SVM are compared. The configurable parameters include forest width, depth and splitting criterion, and are tuned using 30 iterations [38]. Increasing the tree complexity above 32 trees with 128 splits provided no improvement in performance but did increase computation time. The classification process is adjusted using SC and FLDA to find both the peak attainable results and their optimal efficiency.

### A. Device Classification using feature selection methods

In this section the classification results using two different feature selection techniques are presented. The aim is to reduce the device observation period, while maintaining a high degree of accuracy. In all experiments the Decision Forest is configured to 32 trees with 128 splits per node. For validation k-fold cross validation is adopted. The performance of each model is measured using AUC, Sensitivity and Specificity. Tables 3 and 4 highlight the results for both the Decision Forest and SVM using FLDA.

TABLE 3
DECISION FOREST USING FLDA

| Device | Sensitivity | Specificity | AUC |
|---|---|---|---|
| Kettle | 98.67 | 96.92 | 97.31 |
| Microwave | 93.33 | 98.08 | 97.02 |
| Toaster | 73.68 | 95.41 | 90.81 |
| Washing Machine | 86.67 | 96.31 | 94.22 |
| Cooker | 81.33 | 94.98 | 92.09 |

TABLE 4
SVM USING FLDA

| Device | Sensitivity | Specificity | AUC |
|---|---|---|---|
| Kettle | 98.67 | 94.59 | 95.62 |
| Microwave | 88.00 | 98.64 | 95.95 |
| Toaster | 52.63 | 93.49 | 70.18 |
| Washing Machine | 53.33 | 95.69 | 86.06 |
| Cooker | 85.33 | 81.78 | 82.56 |

Tables 5 and 6 present the results using SC. The results show that the Decision Forest obtained slightly better results for the kettle and washing machine class. However, a notable reduction in performance across the reaming classes is observed.

TABLE 5
DECISION FOREST USING SC

| Device | Sensitivity | Specificity | AUC |
|---|---|---|---|
| Kettle | 96.00 | 98.44 | 97.89 |
| Microwave | 93.33 | 98.08 | 93.33 |
| Toaster | 73.68 | 94.06 | 89.78 |
| Washing Machine | 85.33 | 97.03 | 94.48 |
| Cooker | 84.00 | 93.91 | 91.81 |

TABLE 6
SVM USING SC

| Device | Sensitivity | Specificity | AUC |
|---|---|---|---|
| Kettle | 98.67 | 94.2 | 95.32 |
| Microwave | 86.67 | 98.65 | 95.64 |
| Toaster | 51.32 | 92.83 | 83.58 |
| Washing Machine | 60.00 | 94.86 | 89.89 |
| Cooker | 82.67 | 83.83 | 83.58 |

In the case of the washing machine and toaster class, the SVM produced better results when using SC as shown in Table 6. The performance improved for the cooker, toaster and washing machine class with a slight reduction for the kettle and microwave class.

### B. Behavioural Analysis

Behavioural changes in dementia patients often include alterations from normal routine behaviour, for example, sleep disturbances. These abnormalities tend to increase in severity and frequency as dementia progresses. Occasionally a person with dementia will exhibit an increase in certain behaviours in the late afternoon or early evening. This condition is often referred to as sundowning syndrome.

Behavioural alterations such as sundowning syndrome can be detected by detecting gradual changes in energy usage over long observation periods. Figure 10 highlights how data, obtained from the clinical trial, can be used to provide insights about significant behavioural patterns (depicted by the link thickness).

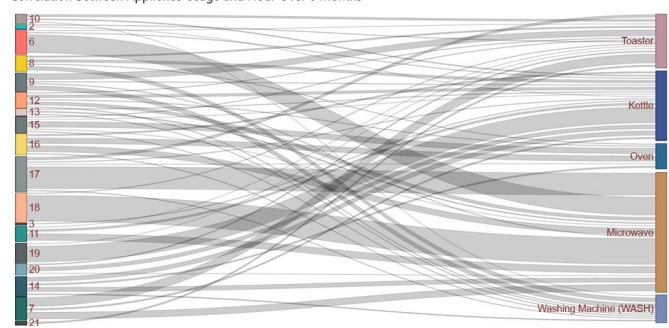

**Fig. 10:** Sankey Plot Showing the Degree of Correlation between Device Usage and Hour

The lines between appliance and time of day emphasis established behavioural routines, therefore alterations in either link proportionality or association provides indicators for disease progression and potential relapses.

Figure 10 allows clinicians to establish the routine behaviour of a patient and define a baseline for anomaly detection, i.e. deviations from what is considered normal behaviour for that patient. During a longer longitudinal clinical trial where the Sankey Plot could be played for a three year period, we would expect to see changes between correlations and their associated strengths.

Again, using data from the clinical trial, Figure 11 shows the results for anomaly detection. The inliers shown in green represent normal appliance interactions. Each cluster represents a specific appliance class. The outliers are depicted in red where both the kettle and toaster classes reside outside the patient's normal routine behaviour. In total three kettles where used on three separate occasions between the hours of 00:00 and 05:00 while a single interaction with a toaster was detected during the same observation period. Throughout the trial a total of 4 sleep disturbances where observed. This may indicate to clinicians important insights into the speed and progression of dementia. Over a longitudinal study, we would expect the green data points to drop lower as a person progresses with their dementia resulting in many more anomalies over time.

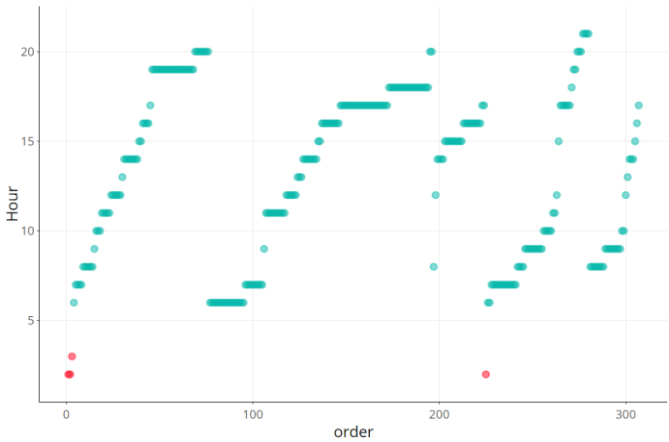

Fig. 11. Sleep Disturbances for Patient 1 using Z-Score Anomaly Detection

## VI. DISCUSSION

The results in this paper demonstrate how the analysis of electricity usage through load disaggregation can be used to model behavioural routines. Machine learning algorithms can sufficiently identify five appliances which include the kettle, microwave, toaster, electric oven and washing machine. The results from the clinical trial show that important ADL's can be detected and used to facilitate behavioural analysis, using these five devices alone. The AUC values across all appliance classes are encouraging and in many cases the random decision forest using FLDA is capable of detecting appliances from aggregate load with very high sensitivity and specificity values.

Participants in the trial included patients with mild to moderate dementia. Data received from the system was used to create a personalised behavioural baseline to continually analyse and detect dementia progression or sudden relapses. The results from the clinical trial show that the system can accurately established patient routines, while detecting anomalies in behaviour. Z-score anomaly detection values show high sensitivly to abnormal device interactions (4 sleep disturbances during the 6-month trial). These results are again encouraging and warrant further investigation.

We believe that as patients progress with dementia or exhibit periods of relapse, an increase in outlier detection will be evident where normal behaviour (show in Figure 11) slowly begins to drop into regions of abnormal behaviour. Or in the case of Figure 10, the correlations between time of day and appliance type begin to shift and the thickness of connecting lines either shrink or become thicker over time. The detection of anomalies in this way will facilitate early intervention and provide a never-before-seen objective measure for dementia progression with near real-time detection. If successful this will allow care packages or enhanced support to be provided as and when required, but more importantly, much earlier thus mitigating the effects of crisis point care.

## VII. CONCLUSION AND FUTURE WORK

The approach in this paper presents a foundational, novel technology, to monitor dementia patients and disease progression at home. The solution is different from any other telehealth offering whereby patients are not required to interact with technology or learn anything new to benefit from the services the system provides. Using the data collected from electricity readings the technology can accurately identify the use of individual electrical devices in the home and the routine behaviours of people to detect when anomalies occur. This novel approach facilitates the detection of specific ADLs in ways that has never been possible before without incurring considerable costs or overtaxing the cognitive needs of the patient.

Therefore, the system contributes significantly to the field of Ambient Assistive living (AAL). In the UK, the effects of an ever-ageing population are becoming increasingly harder to manage. Consequently, a variety of challenges for both health and social care providers have been introduced. Using this foundational technology will help deliver never before seen services to a variety of medical domains and support patients in ways that were not possible a few years ago. Proposed applications include issuing alerts to carers when unusual home appliance activity patterns are recognised. This will provide an automated way of identifying significant events such as sleep disturbances, inactivity and condition monitoring (to inform treatment needs). The ability to identify appliance usage patterns will help to facilitate a greater understanding of patients living at home on their own with life limiting medical conditions. In this way, signs of worrying behaviour or events in a person's life can be detected much earlier and interventions can be administered much quicker.

While the clinical trial has helped us to understand some initial parameters of smart meter home care monitoring there still remains a great deal of work to be done. In order to understand how the usage of appliances and the consumption of energy changes for dementia patients over time the research team are planning to undertake a 2.5 year longitudinal study in partnership with Mersey Care NHS Foundation Trust. In total 50 patients will be recruited where they have a diagnosis of mild to moderate stage dementia of any type. During this time energy usage will be collected and analysed and the results will be compared with regular patient assessment scores. The functional ability of each patient in the study will be assessed every 6 months using the Bristol Activity of Daily Living score and also when abnormal behaviours are detected and deemed important to warrant clinical assessment. Cognitive functions will be monitored using the MMSE and compared with device usage patterns for a set of given observation periods. The study will determine if any correlation between the two metrics exists and if so to what extent. Examining if there is a change in the distribution of energy usage could provide a mechanism for tracking the progression of dementia. During the study a set of clinical markers will be established based on the correlation between appliance interactions and the associated cognitive score. These markers will act as a trigger for the deployment of appropriate interventions such as the introduction of care packages. The markers will be used in further case-control clinical trials to measure the overall effectiveness of the interventions.

## ACKNOWLEDGMENTS

The authors of this paper would like to thank Professor Chris Dowrick for his advice and support in the early stages of the project. Thanks are also given to Pauline Parker, Head of Research and Jill Pendleton, from Community Services at Mersey Care NHS Foundation Trust. The authors would like to sincerely thank both patients who participated in the trial. Their advice and willingness to help future patients through clinical research is inspirational.

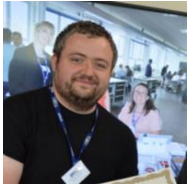
**Dr Carl Chalmers** is a Senior Lecturer in the Department of Computer Science at Liverpool John Moores University. Dr Chalmers's main research interests include the advanced metering infrastructure, smart technologies, ambient assistive living, machine learning, high performance computing, cloud computing and data visualisation. His current research area focuses on remote patient monitoring and ICT-based healthcare. He is currently leading a three-year project on smart energy data and dementia in collaboration with Mersey Care NHS Trust. The current trail involves monitoring and modelling the behaviour of dementia patients to facilitate safe independent living. In addition he is also working in the area of high performance computing and cloud computing to support and improve existing machine learning approaches, while facilitating application integration.

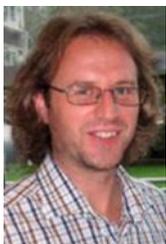
**Dr Paul Fergus** is a Reader (Associate Professor) in Machine Learning. Dr Fergus's main research interests include machine learning for detecting and predicting preterm births. He is also interested in the detection of foetal hypoxia, electroencephalogram seizure classification and bioinformatics (polygenetic obesity, Type II diabetes and multiple sclerosis). He is also currently conducting research with Mersey Care NHS Foundation Trust looking on the use of smart meters to detect activities of daily living in people living alone with Dementia by monitoring the use of home appliances to model habitual behaviors for early intervention practices and safe independent living at home.

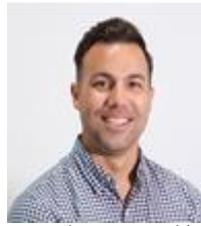
**Dr Casimiro Curbelo Montañez** is a Research Assistant in the Department of Computer Science at Liverpool John Moores University (LJMU), UK. Dr Curbelo received his B.Eng. in Telecommunications in 2011 from Alfonso X el Sabio University, Madrid (Spain). In 2014, he obtained an MSc in Wireless and Mobile Computing from LJMU. He completed his PhD in 2019 also at LJMU. He is currently part of a research team, working on strategies to detect anomalies utilising the smart metering infrastructure. His research interests include various aspects of data science, machine learning and their applications in Bioinformatics.

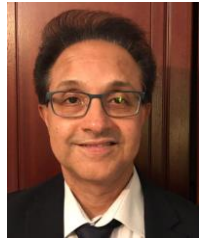
**Dr Sudip Sikdar** is a consultant in old age psychiatry and has worked in Liverpool since 2001. He did his undergraduate training as a doctor in Calcutta, and his post graduate psychiatry training the famous Post Graduate Institute of Medical Education and Research in Chandigarh in India. Dr Sikdar did further post graduate training in psychiatry in UK followed by higher specialist training in old age psychiatry before being appointed as a consultant.

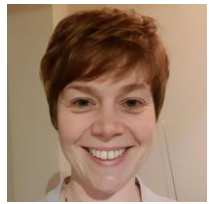
**Dr Freya Ball** graduated from the University of Liverpool in 2010. She is currently an ST6 in old age psychiatry training with the North West Deanery in the Merseyside region.

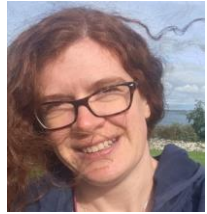
**Dr Bryony Kendall** is a GP partner at Aintree Park Group Practice in Liverpool. She has a special interest in safeguarding, and in the usage of machine learning for practical, sustainable and evidence based interventions to aid the most vulnerable.